# Spatial phase coherence in femtosecond coherent Raman scattering

Ali Hosseinnia,[1,*] Michele Marrocco,[2,†] Francesco Vergari,[2,3] Meena Raveesh,[4] Sebastian Riewer,[1] Ashutosh Jena,[1] Abhishek Kushwaha,[1] Francesco Mazza,[1] Mark Linne,[5] Joakim Bood,[4] and Isaac Boxx[1]

[1]Chair of Optical Diagnostics in Energy, Process and Chemical Engineering, RWTH Aachen University, 52062 Aachen, Germany
[2]ENEA, via Anguillarese 301, Roma I-00123, Italy
[3]Dipartimento di Fisica, Università di Roma "La Sapienza", Roma I-00185, Italy
[4]Division of Combustion Physics, Department of Physics, Lund University, S-221 00 Lund, Sweden
[5]School of Engineering, The University of Edinburgh, Edinburgh EH8 3JL, Scotland, UK

Conventional femtosecond coherent laser spectroscopy predominantly focuses on the temporal phase coherence through time- or frequency-resolved methods. In this work, we suggest an alternative experimental framework based on spatial phase coherence. The intrinsic spectral dispersion of wavevectors in femtosecond pulses and sample dimensions exceeding the laser wavelength create a compelling basis to establish spatial phase coherence as a novel and robust foundation for femtosecond laser spectroscopy. Using rotational Raman coherence in gas molecules as a case study, we analyze the transverse spatial distribution of the third-order signal generated by a rotational wave-packet. Our findings reveal apparent temporal shifts and distortions in time-resolved signals that arise in conventional measurements lacking sensitivity to spatial phase coherence. Moreover, we demonstrate that spatial phase coherence can serve as a useful tool for thermometric applications, showcasing its sensitivity to temperature variations. These discoveries open new avenues in femtosecond laser spectroscopy, including an alternative single-shot detection scheme, a new form of Raman coherence imaging and molecular species quantification during overlapping fractional revivals.

Phase coherence is a cornerstone of collective phenomena, underpinning many remarkable advancements in both classical and quantum physics [1-7]. In laser spectroscopy, phase coherence is used in time- and frequency-resolved approaches or their combination, enabling insights into applications of atomic, molecular and optical physics [8-11]. Beyond these conventional schemes, phase coherence is more comprehensively described in the framework of space-time four-vectors $(ct, \boldsymbol{r})$ and their reciprocal space duals $(\omega/c, -\boldsymbol{k})$. The more general context dictates that the temporal phase term $\omega t$ is complemented by the spatial term $-\boldsymbol{k} \cdot \boldsymbol{r}$. The link between spatial coordinates and wavevectors highlights the broader utility of phase coherence, extending its applications beyond traditional time- or/and frequency-resolved methods of laser spectroscopy. Examples are known in other branches of physics, such as stellar interferometry [12] or Bose-Einstein condensation [13]. From this perspective, spatial phase coherence offers a complementary avenue for spectroscopic investigations.

In this study, we investigate experimentally the spatial component of the signal's phase in relation to its temporal counterpart. Our aim is to explore the potential of space-wavevector coupling as a foundation for a new experimental platform designed to achieve a deeper understanding of phase invariance in ultrafast coherent laser spectroscopies. Additionally, we seek to define the influence of the spatial phase component on temporal phase information.

This research employs the framework of coherent Raman scattering (CRS) [9-11, 14-19]. In this context, physical information is routinely gained under the assumption of localized interactions, as described by conventional response theories based on either third-order susceptibility [9, 10] or anisotropic refractive index contributions [20] arising from molecular alignment [21].

The space-dependent interactions of this work are studied by means of a simplified version of the standard experimental set-up designed to study ultrafast Raman coherence in gases [16, 17, 22, 23]. In particular, the set-up has been streamlined to enhance practicality and detection of spatial phase effects. The simplification amounts to the use of one single femtosecond laser source and the replacement of the spectroscopic detection tools with an imaging system. The femtosecond pulses are thus degenerate but spectrally broad, with the two pump (or driving) electric fields derived from the same laser source, providing two distinct frequencies within its spectral bandwidth. A third laser field serves as the probe, reading the molecular coherence after a controlled delay. The key details of the experimental setup are outlined below.

---

*Contact author: hosseinnia@lom.rwth-aachen.de
†Contact author: michele.marrocco@enea.it

The driving and reading pulses were generated by a Ti:Sapphire laser system (Coherent Hidra amplifier pumped by a Mantis oscillator), delivering transform-limited 50 fs pulses at 800 nm with spatial profile close to a Gaussian profile of 17 mm. The laser system could deliver pulses with 25 mJ at a repetition rate of 10 Hz. They were split into two parts of equal energy using a beam splitter. One of these was further divided by a secondary beam splitter with 20% transmission. The reflected portion was directed through another beam splitter, producing two synchronous excitation pulses driving the Raman coherence. The transmitted pulse (probe) was routed through a translation stage to probe Raman coherence at specific time delays. All pulses were focused using a lens with a 55-cm focal length in a three-dimensional phase-matching configuration as shown in Fig. 1(a) (folded BOXCARS).

The pulse energies, varied within the range of tens of μJ, were used to generate Raman coherence in air under ambient pressure conditions and at temperatures up to 600 °C. A heat gun and a K-type thermocouple were used for thermometric control. Signal collimation was achieved using a lens with a 20-cm focal length. The three-dimensional pulse geometry effectively isolated the CRS signal within the cone illuminated by the phase-matched radiation from nitrogen and oxygen molecules.

The signal was captured using a thermoelectrically cooled, back-illuminated CCD camera (Newton 940, Andor-Oxford Instruments) at the distance of approximately 60 cm from the probe volume. Selecting a restricted pixel area on the camera allowed the coherent signal evolution to be tracked as the delay progressed, mimicking conventional time-resolved approaches [17, 22-28]. To achieve a recording frame rate of 10 Hz that matched the laser repetition rate, an 8×8 binning was performed on the CCD chip. Conventional time-resolved measurements were conducted by isolating a small region of the CCD chip. An example is shown in Fig. 1(b). The collapses and revivals of the rotational wave packet were distinctly observed at the expected delays, corresponding to fractional values of the full revival time of $1/(2Bc)$ where $B$ is the rotational constant [29]. Two sets of fractional revivals are thus recognizable in Fig. 1(b). For nitrogen, the full revival occurs at approximately $\tau_{N_2} \cong 8.38\ ps$ and, reflecting the rotational dynamics during the free evolution of the Raman coherence, fractional revivals at $\tau_{N_2}/4 \cong 2.09\ ps$, $\tau_{N_2}/2 \cong 4.19\ ps$, $3\tau_{N_2}/4 \cong 6.28\ ps$ are also detected. Similarly, for oxygen, the full revival occurs at $\tau_{O_2} \cong 11.59\ ps$ with fractional revivals appearing at $\tau_{O_2}/4 \cong 2.90\ ps$, $\tau_{O_2}/2 \cong 5.80\ ps$, and $3\tau_{O_2}/4 \cong 8.69\ ps$. The pronounced peaks around 8.5 ps are a distinctive feature of Fig. 1(b). They arise from the partial interference between the full revival of nitrogen and the $3\tau_{O_2}/4$ revival of oxygen. Additionally, other fractional revivals of nitrogen for delays exceeding $\tau_{N_2}$ are also shown.

As discussed earlier, measuring the time trace captures the temporal component of phase coherence while keeping the spatial phase fixed. This behavior is illustrated in Fig. 2(a), where the trace represents the signal recorded at a specific pixel of about 100x100 μm$^2$ on the CCD (indicated by the white spot in Fig. 2(b)). The variation of the relative angles of the linear polarization states of the laser pulses was of no consequence, except for a general decrease in signal intensity. The theoretical reconstruction of the time trace is also shown in Fig. 2(a), with details of the simulation method available elsewhere [17].

The comparison between the highlighted region of Fig. 2(a) and the image of Fig. 2(b) recorded at ~4.19 ps delay reveals that the spatial phase serves as a carrier of both past and future time-resolved spectroscopic information. This characteristic underscores the potential of spatial phase coherence to facilitate single-shot measurements, which are traditionally accomplished at a fixed delay using more intricate experimental setups, such as the chirped-probe technique [30] or other phase-engineered approaches [28].

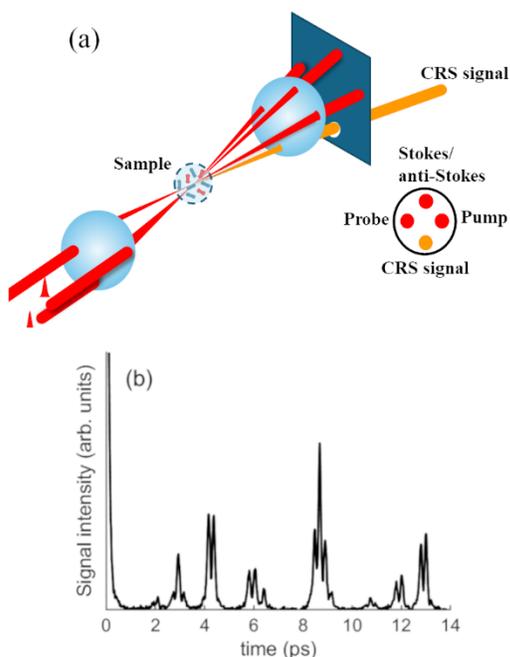

FIG 1. (a) Three-dimensional phase-matching geometry and (b) typical detected time trace as for a conventional time-resolved approach.

Figure 2(c) presents a theoretical reconstruction of the spatial signal generated under the Gaussian approximation for the temporal and spatial profiles of the laser pulses, revealing several key features that are explained next. Most notably, the spots appear elongated at an angle of approximately 70 degrees, reflecting the symmetry of the wavevector components along the horizontal (x-axis) and vertical (y-axis) directions of the CCD plane. The Gaussian structure is projected onto the elliptical shape of the illuminated region caused by the phase-matching geometry. The distinctive features of spatial phase measurements are demonstrated in the video provided by the authors on request. The video showcases a sequence of images, similar to Fig. 2(b), captured at larger time delays. These images replicate the structure of conventional time traces while offering more detailed insights into the evolution of the phase in both its temporal and spatial components. This further highlights the technique's potential for enabling single-shot measurements.

In addition to mapping the time-resolved signal onto one spatial axis of the CCD, another key phenomenon that links spatial and temporal phase components is the intensity swing observed within the same revival. This behavior is captured in Fig. 3, where two experimental time frames of the half revival of nitrogen are compared in Figs. 3(a) and 3(b). The images are collected at two distinct delays whose time separation is 100 fs. The transfer of the signal maximum from one peak to the other is clearly recorded. The simulated images in Figs. 3(c) and 3(d) successfully replicate this phenomenon, providing further evidence of the dynamic relationship between spatial and time coordinates. The intensity swing not only underscores this interplay but also emphasizes the sensitivity of time-resolved measurements to the specific spatial region being observed.

The sensitivity to the spatial phase was further examined by analyzing signals from various regions within the CCD photoactive plane. Fig. 4 provides a comprehensive illustration of this sensitivity to the choice of regions where the time-resolved signal is measured across increasing delays. Below, we limit

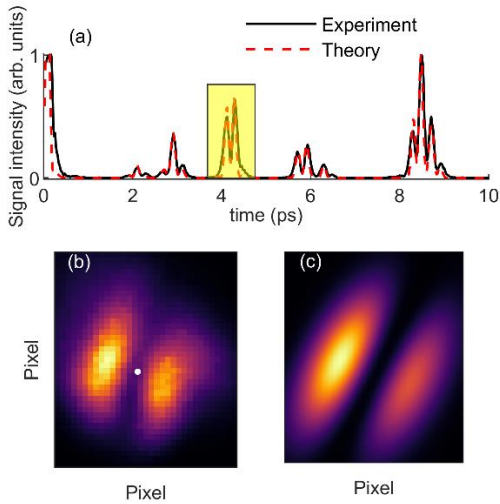

FIG 2. (a) Experimental time trace (solid black line) and corresponding simulation (dashed red line). The experimental data refer to the selected pixel of the image in (b) and highlighted as a white dot. (b) Spatial signal observed at the time delay of about 4.19 ps. The horizontal axis is made of 33 pixels that correspond to the time window of the yellow highlighted region of the time trace in (a). Pixels before the center link to the time before the selected delay. Pixels after the center link to the future of the time trace. (c) Simulation of the experimental image in (b). Details of the simulations are available elsewhere [17].

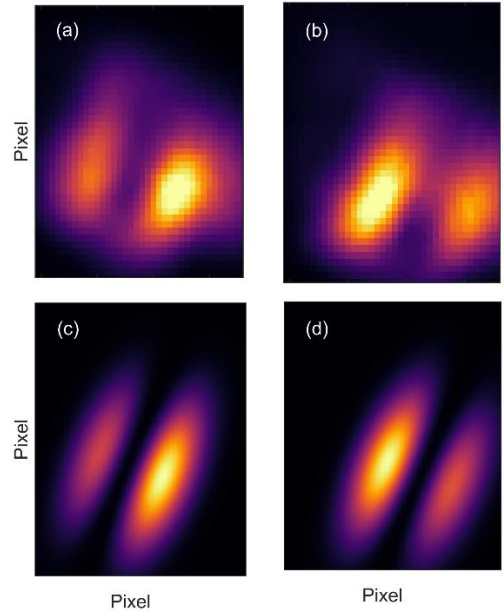

FIG 3. Experimental images (a) and (b) of the nitrogen half revival captured at two delays separated by 100 fs. The more intense signal swings from one peak to the other as the delay increases or decreases. The simulation in the corresponding images (c) and (d) confirms the phenomenon.

ourselves to examine the case of very small regions (single pixels).

The experimental time traces in Fig. 4(a) represent measurements taken from two specific pixels, labeled as points A and B in the image shown in Fig. 4(b). This image reproduces the spatial distribution of the half revival of nitrogen when the delay was 4.16 ps. When an alternative observation point, such as pixel B in Fig. 4(b), is selected, a clear temporal phase shift appears in the corresponding time trace. This is evident in the comparison of experimental traces for points A and B in Fig. 4(a). A simulated time trace that is independent from spatial effects is also shown in Fig. 4(a) as a solid black curve. The temporal discrepancy between the experimental traces arises from the change in the spatial phase due to the spatial motion of the signal as a function of probe delay time. As shown in the video available on request to the authors, the signal traverses the CCD chip, entering from the top left and exiting through the bottom right.

These findings have important implications for molecular alignment studies [21, 26], particularly for manipulating and controlling the rotational states of molecules [31]. Such studies often rely on tightly focusing the signal onto a photodetector, which typically overlooks the temporal and spatial phase differences demonstrated in Fig. 4. Accounting for these effects could enhance the precision and understanding of molecular alignment experiments. Furthermore, the challenges in accurately determining the probe delay may stem from uncertainties related to the spatial selection of the observation region. This issue is frequently avoided by treating the delay as a fitting parameter, allowing it to vary freely over a range of several hundred femtoseconds [28, 30].

An additional application of the proposed CRS technique is thermometry [10, 16, 17, 23, 25-28, 30, 32]. In Fig. 5, we demonstrate the relevance of spatial phase for thermometric measurements performed on a single laser-shot basis. Focusing on the half-revival of nitrogen, Fig. 5(a) shows how the relative intensity between the two peaks of the revival changes when measured in a thermometrically controlled volume of air. As the temperature increases, a well-defined variation in the relative line intensities is observed: fixing one peak as a reference, the intensity of the other decreases progressively with rising temperature. These data were extracted from the average between the horizontal lines in images similar to the one displayed in Fig. 5(b).

The experimental results in Fig. 5(a) are compared with simulations in Figs. 5(c) and 5(d), showing good agreement between theory and experiment. The theoretical analysis confirms that increasing temperature leads to variations in birefringence, generating small changes in the two refractive indices. These changes cause subtle shifts in the spatial phase localization on the CCD chip. Additionally, higher temperatures excite rotational states corresponding to higher J-levels, altering the population density and introducing corrections such as centrifugal distortion to the rotational energy spectrum. The interplay of these effects (variations in birefringence, spatial phase shifts, and changes in rotational level populations) is effectively captured in Figs. 5(c) and 5(d). In particular, Fig. 5(c) is the outcome of linear cuts through the 2D density plot of the simulated signal in the spatial domain. Figure 5(d) depicts instead the simulated time-resolved signal. To achieve this result, a small contribution acting as a background from a permanent molecular alignment [33] induced by the lasers is included in the simulation. The strength of this alignment is calibrated using data recorded at room temperature $T_0$. Subsequently, the scaling factor $T_0/T$ is applied for simulations at other temperatures.

Interestingly, conclusive and repeatable experimental evidence revealed that the strength of the induced permanent alignment in the signal's time trace varies depending on the spatial position from which the time trace is recorded. In the end, the results in Fig. 5 hold promise for both spatially and temporally resolved CRS for precise, temperature-sensitive

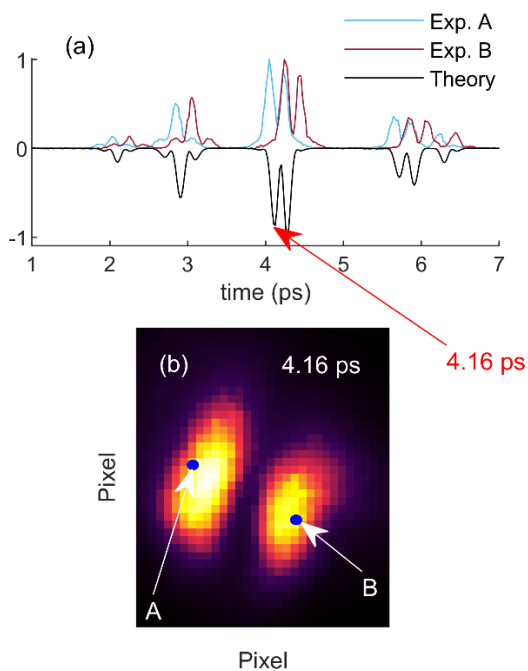

FIG 4. (a) Experimental time traces (blue and red lines) measured at the selected pixels A and B in the image. The theoretical trace (black line) is also plotted. (b) Spatial distribution of the signal measured when the delay was 4.16 ps.

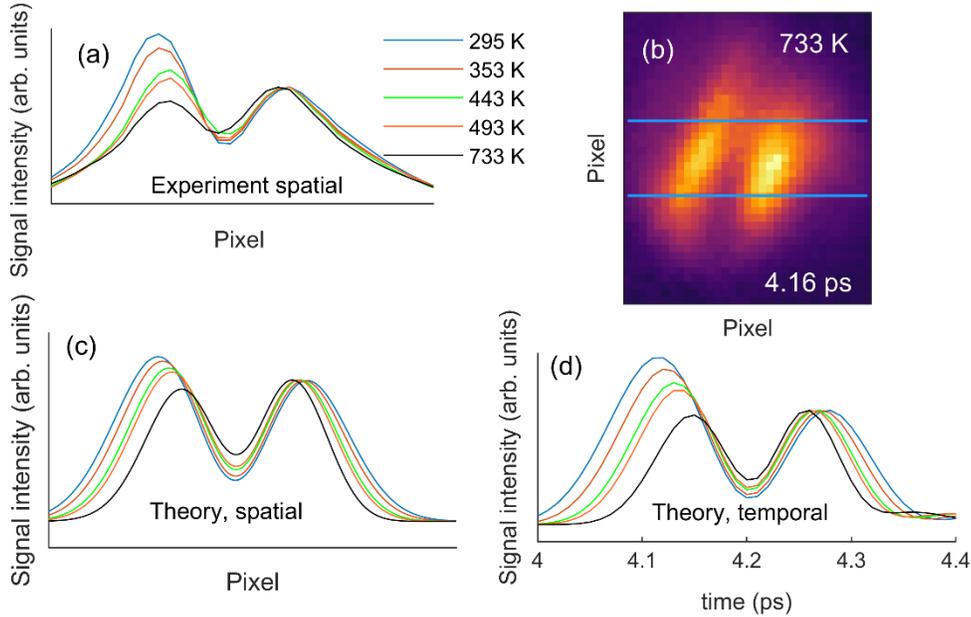

FIG 5. (a) Averaged line cuts extracted from experimental images recorded at different temperatures. One such image, recorded at 733 K, is shown in (b). The averages leading to result in (a) are made over the columns of the pixel matrix defined by the region enclosed between the blue lines visualized in (b) only for the highest temperature. (c) The corresponding simulation of (a) in space domain. (d) Time-domain simulation including an excess of permanent laser-induced alignment normalized to the experimental data at $T_0 = 295\ K$.

diagnostics in gaseous systems with the further potential of single-shot measurements.

In conclusion, we have demonstrated the intimate relationship between the spatial and temporal components of the optical phase, arising from the phase invariance of its four-vector nature. Using Raman coherence of gaseous molecules interacting with femtosecond laser pulses as a case study, we identified limitations in the accuracy of conventional time-resolved measurements caused by the interplay between temporal and spatial phases. Our findings emphasize that a complete understanding of Raman coherence experiments must incorporate spatial information.

Notably, we observed a novel phenomenon, the "intensity swing," which involves a temporal shift of the maximum third-order response between peaks within the same revival. Recognizing this effect could significantly improve the precision of Raman coherence measurements. Additionally, we demonstrated sensitivity to temperature variations, highlighting the potential for single-shot thermometry.

This work paves the way for new techniques in femtosecond laser spectroscopy, offering several promising avenues. These include the development of a new type of Raman coherence imaging and the capability to quantify molecular species by means of the interference arising from overlapping fractional revivals in single-shot recordings. These advancements and their implications will be addressed in a forthcoming publication.

A final comment regards the applicative extensions to vibrational CRS spectroscopy that is common for heavy gas molecules [34] or molecular constituents in soft and condensed matter [35, 36]. For the latter, typical dephasing times are on the order of a few picoseconds, and advanced spectroscopic techniques are employed. These include temporal phase engineering to suppress the non-resonant background [37] or finding a balance between resolution, signal strength, and background interference [15]. Exploring spatial phase effects presents an intriguing alternative, provided the time-modulated signal exhibits sufficient contrast. This consideration is also relevant for single-molecule vibrational studies [38].

*Acknowledgements* – This study was funded by the Deutsche Forschungsgemeinschaft (DFG, German Research Foundation) under Germany´s Excellence Strategy – Cluster of Excellence 2186, The Fuel Science Center" – ID: 390919832, and also was performed within the Italian project "RICERCA E SVILUPPO DI TECNOLOGIE PER LA FILIERA DELL'IDROGENO POR-H2" ("Research and Development of Technologies for Hydrogen Chain", POR-H2, WP2, LA2.2.5), funded by the European Union (NextGeneraionEU) through the Italian Ministry of Environment and Energy Security, under the National Recovery and Resilience Plan (PNRR,